\documentstyle[prl,aps,floats,twocolumn]{revtex}

\input epsf.tex

\newcommand{\postscript}[2]{\setlength{\epsfxsize}{#2\hsize}
   \centerline{\epsfbox{#1}}}

\newcommand{\gev}{\text{GeV}}

\newcommand{\g}{\text{g}}
\newcommand{\cm}{\text{cm}}
\newcommand{\m}{\text{m}}
\newcommand{\km}{\text{km}}
\newcommand{\s}{\text{s}}

\newcommand{\sr}{\text{sr}}
\newcommand{\rearth}{R_{\oplus}}

\newcommand{\be}{\begin{equation}}
\newcommand{\ee}{\end{equation}}
\newcommand{\eg}{{\em e.g.}}

\newcommand{\etal}{{\em et al.}}
\newcommand{\ibid}{{\em ibid.}}
\newcommand{\eqref}[1]{Eq.~(\ref{#1})}

\begin{document}

\draft

\renewcommand{\thefootnote}{\fnsymbol{footnote}}
\setcounter{footnote}{0}

\preprint{
\noindent
\hfill
\begin{minipage}[t]{3in}
\begin{flushright}
MIT--CTP--3122\\
MIT--LNS--01--294\\
hep-ph/0105067
\end{flushright}
\end{minipage}
}

\twocolumn[\hsize\textwidth\columnwidth\hsize\csname
@twocolumnfalse\endcsname

\title{
Observability of Earth-skimming Ultra-high Energy Neutrinos 
}

\author{
Jonathan L.~Feng$^{ab}$
,
Peter Fisher$^c$,
Frank Wilczek$^a$,
Terri M.~Yu$^c$
\vspace*{0.1in}
}

\address{
  ${}^{a}$Center for Theoretical Physics,
  Massachusetts Institute of Technology,
  Cambridge, MA 02139 USA \\
  ${}^{b}$Department of Physics and Astronomy,
  University of California, Irvine,
  Irvine, CA 92697 USA\\
  ${}^{c}$Department of Physics, 
  Massachusetts Institute of Technology,
  Cambridge, MA 02139 USA
}


\maketitle

\begin{abstract}
Neutrinos with energies above $10^8~\gev$ are expected from cosmic ray
interactions with the microwave background and are predicted in many
speculative models.  Such energetic neutrinos are difficult to detect,
as they are shadowed by the Earth, but rarely interact in the
atmosphere.  Here we propose a novel detection strategy:
Earth-skimming neutrinos convert to charged leptons that escape the
Earth, and these leptons are detected in ground level fluorescence
detectors.  With the existing HiRes detector, neutrinos from some
proposed sources are marginally detectable, and improvements of two
orders of magnitude are possible at the proposed Telescope Array.
\end{abstract}





]

\renewcommand{\thefootnote}{\arabic{footnote}}
\setcounter{footnote}{0}

Cosmic neutrinos with energies above $10^8~\gev$, so far unobserved,
have great potential as probes of astrophysics and particle physics
phenomena.  They escape from dense regions of matter and point back to
their sources, thereby providing a unique window into the most violent
events in the universe.  Once they reach the Earth, they interact with
center-of-mass energies far beyond foreseeable man-made colliders and
probe new physics at and beyond the weak scale.

\begin{figure}[tb]
\postscript{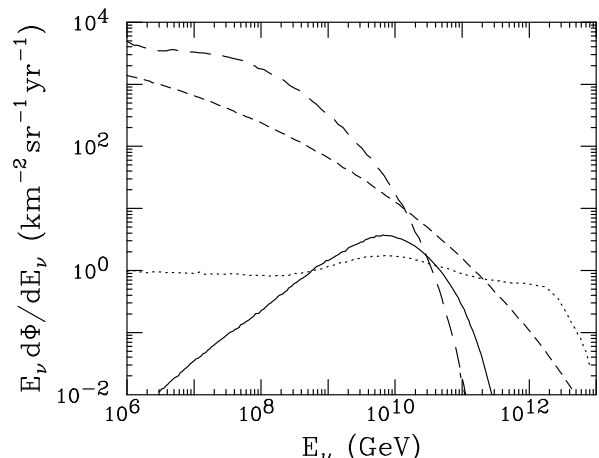}{0.90}
\caption
{Differential fluxes of muon neutrinos ($\nu_{\mu} + \bar{\nu}_{\mu}$)
from Greisen photoproduction~\protect\cite{Stecker:1979ah} (solid),
active galactic nuclei~\protect\cite{Mannheim:1995mm} (long dashed),
topological defects~\protect\cite{Sigl:1999vz} (short dashed), and
$Z$-bursts~\protect\cite{Yoshida:1998it} (dotted).  For maximal
$\nu_{\mu}$--$\nu_{\tau}$ mixing, these fluxes are divided equally
between $\mu$ and $\tau$ neutrinos when they reach the Earth.}
\label{fig:fluxes}
\end{figure}

The sources of ultra-high energy neutrinos range from the
well-established to the highly speculative~\cite{reviews}.  The cosmic
ray spectrum is well-measured up to the Greisen-Zatsepin-Kuz'min (GZK)
cutoff~\cite{Greisen:1966jv,Zatsepin:1966jv} at $5\times
10^{10}~\gev$. Such cosmic rays necessarily interact with the
$2.7^{\circ} \text{K}$ cosmic microwave background through pion
photoproduction $p\gamma \to n \pi^+$, producing ``Greisen neutrinos''
when the pions decay~\cite{Greisen:1966jv,Stecker:1979ah}.  In
addition to this `guaranteed' flux, far larger fluxes are predicted in
models of active galactic nuclei (AGN)~\cite{Stecker:1991vm,%
Mannheim:1995mm} and in proposed explanations of the observed cosmic
rays with energies above the GZK cutoff.  The latter include decays of
topological defects (TDs)~\cite{Hill:1987mn} and
$Z$-bursts~\cite{Weiler:1999sh}.  Fluxes from photoproduction and some
representative hypothesized sources are given in
Fig.~\ref{fig:fluxes}.

The detection of ultra-high energy cosmic neutrinos is, however,
extremely difficult, especially for those with energies above
$10^8~\gev$.  At these energies, the neutrino interaction length is
below $2000~\km$ water equivalent in rock, and so upward-going
neutrinos are typically blocked by the Earth.  This shadowing severely
restricts rates in underground detectors such as
AMANDA/IceCube~\cite{Alvarez-Muniz:2001gb}, which are bounded by
detection volumes of at most $1~\km^3$.  At the same time, the
atmosphere is nearly transparent to these neutrinos.  Even for
quasi-horizontal neutrinos, which traverse an atmospheric depth of up
to $360~\m$ water equivalent, fewer than 1 in $10^3$ convert to
charged leptons.  At the Pierre Auger Observatory, estimated detection
rates are of order 0.1 to 1 event per year for Greisen neutrinos from
the ground array~\cite{Capelle:1998zz,Coutu:1999ub}, with similar
rates for the Auger air fluorescence detectors~\cite{Yoshida:1997ie}.

\begin{figure}[tb]
\postscript{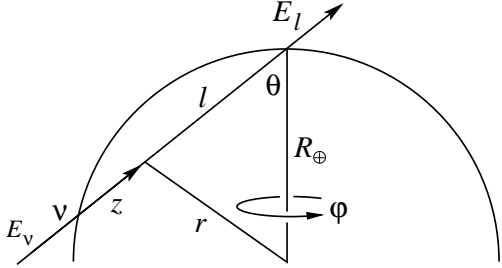}{0.79}
\caption
{A neutrino $\nu$ enters the Earth with energy $E_{\nu}$ at nadir
angle $\theta$ and azimuthal angle $\phi$.  It then travels for
distance $z$ before converting to a charged lepton $\ell$, which exits
the Earth with energy $E_{\ell}$. }
\label{fig:event}
\end{figure}

Here we explore an alternative method for detecting ultra-high energy
neutrinos with $E_{\nu} > 10^8~\gev$.  While upward-going neutrinos
are usually blocked by the Earth, those that skim the Earth, traveling
at low angles along chords with lengths of order their interaction
length, are not.  Some of these neutrinos will convert to charged
leptons.  In particular, muon and tau leptons travel up to ${\cal
O}(10~\km)$ in the Earth at these energies, and so a significant
number of them may exit the Earth and be detected by surface
fluorescence detectors.  A schematic picture of the events we are
considering is given in Fig.~\ref{fig:event}.  This method exploits
both the Earth as a large-volume converter, and the atmosphere as a
large-volume detector.

Upward-going air showers have been discussed
previously~\cite{Domokos}, with an emphasis on differences between
upward-going and conventional showers and the possibility of
space-based detection.  The question of rates was not addressed.  Very
recently, the detection of showers from $\tau$ decays in the Auger
Observatory ground array has been considered~\cite{Bertou:2001vm}.
The possibility of detecting moon-skimming neutrinos through radio
signals has also recently generated
interest~\cite{Alvarez-Muniz:2001rs,Gorham:2001aj}.

Given an isotropic neutrino flux $\Phi_{\nu}$, the resulting
differential flux of charged leptons exiting the Earth is
\begin{equation}
\frac{d\Phi_{\ell} (E_{\ell}, \cos\theta, \phi)}
{dE_{\ell}\, d\cos\theta\, d\phi}
= \frac{1}{2\pi}\!\! \int\!\! 
dE_{\nu} \frac{d\Phi_{\nu} (E_{\nu})}{d E_{\nu}}
K(E_{\nu}, \theta; E_{\ell}) \ ,
\label{conversion}
\end{equation}
where $K$ is the probability that a neutrino entering the Earth with
energy $E_{\nu}$ and nadir angle $\theta$ produces a lepton that exits
the Earth with energy $E_{\ell}$. Such an event requires that (a) the
neutrino survives for some distance $z$ in the Earth, (b) the neutrino
then converts to a lepton, (c) the created lepton exits the Earth
before decaying, and (d) the lepton's energy and position when
produced are such that it leaves the Earth with energy $E_{\ell}$.

The probability for a neutrino with energy $E_{\nu}$ and nadir angle
$\theta$ to survive for a distance $z$ is
\begin{equation}
P_a = \exp \left[ -\int_0^z \frac{dz'}{L^{\nu}_{CC}(E_{\nu}, \theta,
z')} \right] \ ,
\label{Pa}
\end{equation}
where $L^{\nu}_{CC}(E_{\nu}, \theta, z) = [ \sigma^{\nu}_{CC}
(E_{\nu}) \rho (r(\theta, z)) N_A ]^{-1}$ is the charged current
interaction length, with $\sigma^{\nu}_{CC}(E_{\nu})$ the interaction
cross section $\sigma(\nu N \to \ell X)$ for a neutrino with energy
$E_{\nu}$, $\rho(r)$ the Earth's density at distance $r$ from its
center, and $N_A = 6.022 \times 10^{23}~\text{g}^{-1}$.  The distance
$r$ is given by $r^2(\theta, z) = \rearth^2 + z^2 - 2 \rearth z \cos
\theta$, where $\rearth \approx 6371~\km$ is the radius of the Earth.
For $E_{\nu} \agt 10^8~\gev$, the charged current $\nu$ and
$\bar{\nu}$ cross sections are virtually identical, and we may neglect
multiple charged current interactions and neutrino energy degradation
from neutral current processes. Also, at these energies, the optimal
nadir angle for charged lepton production is $90^{\circ} - \theta
\approx 1^{\circ}$.  Leptons produced by Earth-skimming neutrinos
travel essentially horizontally.

The probability for neutrino conversion to a charged lepton in the
interval $[z,z+dz]$ is $dz/L^{\nu}_{CC}(E_{\nu}, \theta, z)$.
However, given that detectable leptons travel nearly horizontally with
path length of ${\cal O}(10~\km)$, this conversion must take place
near the Earth's surface where the Earth's density is $\rho_s =
2.65~\g/\cm^3$. The conversion probability is then well-approximated
by
\begin{equation}
P_b = \frac{dz}{L^{\nu}_{CC\, s}(E_{\nu})} \ ,
\label{Pb}
\end{equation}
where $L^{\nu}_{CC\, s}(E_{\nu}) = [\sigma^{\nu}_{CC}(E_{\nu}) \rho_s
N_A ] ^{-1}$. We assume the lepton takes all of the neutrino energy.
For ultra-high energy neutrinos, the mean inelasticity parameter is
$\langle 1 - E_{\ell}/E_{\nu} \rangle \approx
0.2$~\cite{Gandhi:1996tf}. We therefore expect this assumption to make
only a small difference.

The survival probability $P_c$ for a charged lepton losing energy as
it moves through the Earth is described by the coupled differential
equations
\begin{eqnarray}
dE_{\ell}/dz &=& - \left( \alpha_{\ell} + 
\beta_{\ell} E_{\ell} \right) \rho (r (\theta, z)) \label{Eloss} \\
dP_c/dz &=& - P_c / (c \tau_{\ell} E_{\ell} / m_{\ell}) \ ,
\end{eqnarray}
where $c$ is the speed of light, and $m_{\ell}$ and $\tau_{\ell}$ are
the lepton's rest mass and lifetime, respectively.
Equation~(\ref{Eloss}) parameterizes lepton energy loss through
bremsstrahlung, pair production, and photonuclear interactions, under
the assumption of uniform energy loss.  For the energies of interest
here, $\beta_{\tau} \approx 0.8 \times 10^{-6}~\cm^2/\g$, $\beta_{\mu}
\approx 6.0 \times
10^{-6}~\cm^2/\g$~\cite{Lipari:1991ut,Dutta:2000hh}, and the effects
of $\alpha_{\tau,\mu}$ are negligible.  At the Earth's surface, taus
and muons lose a decade of energy in 11 km and 1.5 km, respectively.
These differential equations are easily solved for a constant density
$\rho_s$, and the survival probability is
\begin{equation}
P_c = \exp \left[ \frac{m_{\ell}}{c \tau_{\ell} \beta_{\ell} \rho_s } 
\left( \frac{1}{E_{\nu}} - \frac{1}{E_{\ell}} \right) \right] \ .
\label{Pc}
\end{equation}
Muon lifetimes are long enough that $P_c \simeq 1$, but this factor
may play a significant role for taus.

Finally, the lepton's energy and location when produced must be
consistent with an exit energy $E_{\ell}$.  {}From \eqref{Eloss}, for
constant density $\rho_s$ and negligible $\alpha_{\ell}$, this
condition is enforced with the delta function
\begin{equation}
P_d = \delta \left( E_{\ell} - E_{\nu} e^{- \beta_{\ell} 
\rho_s \left( 2\rearth\cos\theta - z \right)} \right) \ .
\label{Pd}
\end{equation}

Combining Eqs.~(\ref{Pa}), (\ref{Pb}), (\ref{Pc}), and (\ref{Pd}), the
kernel is then
\begin{equation} 
K(E_{\nu}, \theta; E_{\ell}) = 
\int_0^{2\rearth \cos\theta} P_a P_b P_c P_d \ .
\label{complete}
\end{equation}
However, \eqref{complete} may be further simplified, because the
lepton's range in Earth is far less than the typical neutrino
interaction length.  The kernel is therefore dominated by the
contribution from $z \approx 2\rearth \cos\theta$, and we may replace
$z$ with $2\rearth \cos\theta$ in $P_a$.  The only remaining
$z$-dependence is in $P_d$. Using $\int dz\, \delta( h(z)) = | dh/dz
|^{-1}_{h=0}$, the $z$ integration yields
\begin{eqnarray}
K(E_{\nu}, \theta; E_{\ell}) &\approx&  
\frac{1}{L^{\nu}_{CC\, s}(E_{\nu})}
e^{-\int_0^{2\rearth \cos\theta} 
\frac{dz'}{L^{\nu}_{CC}(E_{\nu}, \theta, z')}} \nonumber \\
&\times& 
\exp \left[ \frac{m_{\ell}}{c\tau_{\ell}\beta_{\ell}\rho_s} 
\left( \frac{1}{E_{\nu}} - \frac{1}{E_{\ell}} \right) \right] 
\frac{1}{E_{\ell} \beta_{\ell} \rho_s } \ .
\label{practical}
\end{eqnarray}
In our calculations, we use the kernel of \eqref{practical} with the
Preliminary Earth density profile~\cite{preliminary}. Our cross
section evaluation closely follows
Refs.~\cite{Gandhi:1996tf,Gandhi:1998ri}; details will be presented
elsewhere~\cite{inprep}.

Muons and taus may be detected in fluorescence detectors either
directly or indirectly through their decay products. We have evaluated
rates for all of these possibilities~\cite{inprep}; here we
concentrate on the most promising signal from electromagnetic energy
in $\tau$-decay showers.  The recent discovery of near-maximal
$\nu_\mu$--$\nu_\tau$ mixing~\cite{Learned:2000qq} implies that, even
at the high energies of interest here, $\nu_\mu:\nu_\tau = 1:1$ at the
Earth's surface.  We assume also that tau decay initiates an
electromagnetic shower with probability $B_{\text{EM}} = 80\%$ and a
typical energy, averaged over all $\tau$ decay modes weighted by
branching fraction, of $E_{\text{EM}} = \frac{1}{3} E_{\tau}$.  At
energies of order $10^{10}~\gev$, the typical shower length is $\sim
10~\km$ in the low atmosphere.

We follow the analysis of Ref.~\cite{Baltrusaitis:1985mx} to estimate
the effective aperture for $\tau$-decay induced showers.  The signal
from an electromagnetic shower must compete with the average noise
from the night sky.  By considering the signal to background ratio in
individual photomultiplier tubes, the energy required for an
electromagnetic shower to be detected was found to be
\begin{equation}
E_{\text{EM}} = E_d R_p^{3/2} e^{R_p/\lambda_R} \ ,
\label{Rp}
\end{equation}
where $R_p$ is the shower's impact parameter in km, and $\lambda_R
\approx 18~\km$ is the Rayleigh scattering
length~\cite{Baltrusaitis:1985mx}.  $E_d$ is an energy characteristic
of the detector.  In particular, $E_d \propto \sqrt{\Delta \theta/D^2}
\propto \sqrt{d/D^3}$, where $\Delta \theta = d/D$ is the angular
acceptance of each photomultiplier tube, and $d$ and $D$ are the
diameters of the photomultiplier tubes and mirror aperture,
respectively.  For Fly's Eye, requiring a 4$\sigma$ triggering
threshold, the value $E_d = 10^8~\gev$ was verified to reproduce the
experimental data well~\cite{Baltrusaitis:1985mx}.  For HiRes, $D$ has
been increased from $1.575~\m$ to $2.0~\m$ and $d$ reduced from
$14.4~\cm$ to $3.5~\cm$~\cite{Abu-Zayyad:2000uu}; we therefore take
$E_d \approx 3.2\times 10^7~\gev$.  For each module of the proposed
Telescope Array, and requiring a 4$\sigma$ signal, $E_d$ has been
estimated to be roughly $4\times 10^5~\gev$~\cite{TAdesignreport}.
Finally, the fluorescence detectors of the Auger
Observatory~\cite{Auger} will also be sensitive to Earth-skimming
events; we expect their sensitivity to lie somewhere between that of
HiRes and Telescope Array.

\begin{figure}[tb]
\postscript{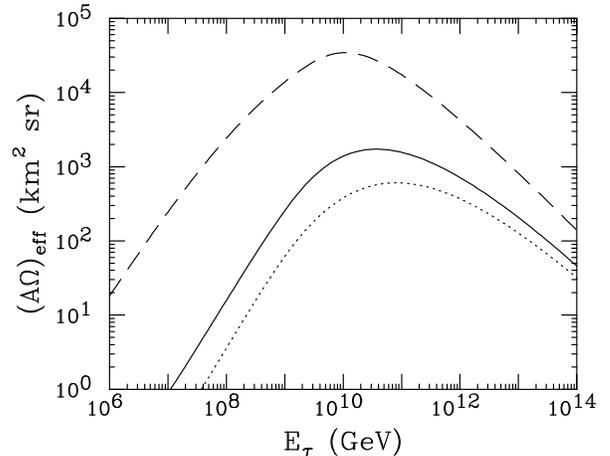}{0.90}
\caption
{Effective aperture estimates for the detection of Earth-skimming
$\tau$ leptons through their decays to electromagnetic showers for
Fly's Eye (dotted), HiRes (one site) (solid), and Telescope Array (one
site) (dashed).}
\label{fig:apertures}
\end{figure}

Following Ref.~\cite{Baltrusaitis:1985mx}, we assume that showers are
detected if and only if initiated within distance $R_p$ of the
detector.  We also make use of the fact that, at these energies, all
$\tau$ leptons exit the Earth horizontally.  Apertures for
Earth-skimming taus for each of the three detectors discussed above
are given in Fig.~\ref{fig:apertures}.  In each case, the aperture
rises with energy until time dilation causes taus to decay too late to
be detected. The HiRes aperture peaks at $2000~\km^2~\sr$ near
$3\times 10^{10}~\gev$.  With increased sensitivity, however, the
aperture peak rises and moves to lower energies, significantly
enhancing detection rates.

Given the kernel function $K(E_{\nu},\theta;E_{\tau})$ and effective
apertures $(A\Omega)_{\text{eff}}(E_{\tau})$, the number of tau
leptons detected is $N_{\tau}\! =\! \int dE_{\nu} dE_{\tau}
d\cos\theta d\phi \cos\theta \frac{d\Phi_{\nu}}{dE_{\nu}} K
(A\Omega)_{\text{eff}} T D$, where $d\Phi_{\nu}(E_{\nu})/dE_{\nu}$ is
the differential flux originating from a given neutrino source, $T$ is
the time an experiment runs, and $D$ is the duty cycle. To account for
the requirement of clear moonless nights for fluorescence detection,
we take $D = 10\%$, corresponding to an observing period of $3 \times
10^6~\s$ per year.

Event rates for the four neutrino sources given in
Fig.~\ref{fig:fluxes}, binned by tau energy, are summarized in
Table~\ref{table:I}.  [Note that these rates are suppressed relative
to those presented in an earlier version of this paper.] We assume 2
and 11 detectors for HiRes and Telescope Array, respectively; some
reduction from overlapping fields of view may be expected.  For HiRes,
we find that neutrinos from AGN and TDs are marginally detectable.
For Telescope Array, these rates are enhanced by more than two orders
of magnitude --- several Greisen neutrinos per year can be detected,
and tens to hundreds of AGN and TD neutrinos are possible.  Note that
the rates may be significantly enhanced by including multi-bang
events, which we have neglected, and also if the $\tau$ energy loss,
dominated by uncertain photo-nuclear interactions, is less than our
conservative assumption.

\renewcommand{\arraystretch}{0.75}
\begin{table}[t!]
\caption{Expected number of $\nu_{\tau}$-induced electromagnetic
showers detected by atmospheric fluorescence.  Three years of running
with duty cycle $D=10\%$ is assumed. \label{table:I} }
\begin{tabular}{lccccc}
Detector & $E_{\tau}$ (GeV)
& \multicolumn{1}{c}{Greisen}
& \multicolumn{1}{c}{AGN}
& \multicolumn{1}{c}{TD}
& \multicolumn{1}{c}{$Z$-burst}
\\ \hline
Fly's Eye
& $10^{8}-10^{9}$    & 0.0039  & 0.051  & 0.0098  & 0.00028 \\
& $10^{9}-10^{10}$   & 0.0021  & 0.027  & 0.012   & 0.0015  \\
& $10^{10}-10^{11}$  & 0.00082 & 0.0011 & 0.0030  & 0.0014  \\
& $10^{11}-10^{12}$  & $-$     & $-$    & 0.00018 & 0.00042 \\
& $10^{12}-10^{13}$  & $-$     & $-$    & $-$     & $-$     \\
& {\em Total}        & 0.0068  & 0.079  & 0.025   & 0.0036  \\
\hline HiRes
& $10^{8}-10^{9}$    & 0.0033  & 0.43   & 0.083   & 0.0024  \\
& $10^{9}-10^{10}$   & 0.017   & 0.22   & 0.094   & 0.012   \\
& $10^{10}-10^{11}$  & 0.0055  & 0.0077 & 0.020   & 0.0092  \\
& $10^{11}-10^{12}$  & $-$     & $-$    & 0.00086 & 0.0019  \\
& $10^{12}-10^{13}$  & $-$     & $-$    & $-$     & 0.00011 \\
& {\em Total}        & 0.026   & 0.66   & 0.20    & 0.026   \\
\hline Telescope
& $10^{8}-10^{9}$    & 1.4     & 230    & 41      & 1.0     \\
Array
& $10^{9}-10^{10}$   & 3.2     & 50     & 20      & 2.3     \\
& $10^{10}-10^{11}$  & 0.62    & 0.93   & 2.2     & 0.93    \\
& $10^{11}-10^{12}$  & $-$     & $-$    & 0.045   & 0.094   \\
& $10^{12}-10^{13}$  & $-$     & $-$    & 0.00035 & 0.0033  \\
& {\em Total}        & 5.2     & 280    & 63      & 4.3     \\
\end{tabular}
\end{table}

Hundreds or even tens of events will shed light on many aspects of
ultra-high energy astrophysics.  The energy spectrum of detected
events varies from source to source, as evident in
Table~\ref{table:I}.  With many events, the source energy spectrum may
be determined by deconvolving the observed spectrum with the kernel
function.  Note also that these rates may be improved with detectors
that cover the sky densely very near the horizon or by filters
optimized for nearly horizontal events.  Placement of detectors in
valleys, which effectively enhances the conversion volume, may also
improve detection rates.

Earth-skimming neutrinos also open up other possibilities for
detection.  Cerenkov radiation provides an alternative signal for
showers initiated by $\tau$ decay.  Conventional air shower arrays,
which deploy a large number of modules over a horizontal area, are not
optimally adapted to Earth-skimming events.  It is interesting to
contemplate `vertical' arrays, say on the side of a mountain, that
would intercept Earth-skimming showers originating from a very large
surrounding area. Earth-skimming events may also be detected from
space, as in the OWL/Airwatch proposal~\cite{OWL}.

\noindent {\em Acknowledgements}. We thank F.~Halzen for helpful
discussions and for bringing Ref.~\cite{Bertou:2001vm} to our
attention, and A.~Kusenko and T.~Weiler for helpful correspondence.
JLF also thanks E.~Kearns, J.~Rosner, and C.~Walter for conversations
about future experiments. This work was supported in part by the
U.~S.~Department of Energy under cooperative research agreement
DF--FC02--94ER40818.

\end{document}